\documentclass{iopart}
\usepackage[T1]{fontenc}
\usepackage{graphicx}
\usepackage{url,graphics,epsfig}
\usepackage{amssymb}

\begin{document}
\setlength{\arraycolsep}{2.5pt}             
\jl{2}
%
%
%
\def\etal{{\it et al~}}
\def\newblock{\hskip .11em plus .33em minus .07em}
%
%
%
%
%
%
\title[Photoionization of  Se$^+$ ions]{Photoionization Cross-Sections for  the trans-iron element Se$^+$ from 18 eV to 31 eV}

 \author{B M McLaughlin$^{1,2}$\footnote[1]{Corresponding author, E-mail: b.mclaughlin@qub.ac.uk}
                and C P Ballance$^{3}$\footnote[2]{Corresponding author, E-mail: ballance@physics.auburn.edu}}

\address{$^{1}$Centre for Theoretical Atomic, Molecular and Optical Physics,
                            School of Mathematics and Physics,
                            The David Bates Building, 7 College Park, 
                            Queen's University Belfast, Belfast BT7 1NN, UK}

\address{$^{2}$Institute for Theoretical Atomic and Molecular Physics,
                            Harvard Smithsonian Center for Astrophysics, 
                            Cambridge, MA 02138, USA}

\address{$^{3}$Department of Physics,  206 Allison Laboratory,
                            Auburn University, Auburn, AL 36849, USA}

\begin{abstract}
Absolute photoionization cross-section calculations are presented for Se$^+$ 
using large-scale close-coupling calculations within the 
Breit-Pauli and Dirac-Coulomb R-matrix approximations.  
The results from our theoretical work are compared with recent 
measurements \cite{Esteves2010,sterling11,Esteves11} made at the Advanced Light Source (ALS) 
radiation facility in Berkeley, California, USA.
We report on results for the photon energy range 18.0~eV -- 31.0~eV,
 which spans the ionization thresholds of the $\rm ^4S^o_{3/2}$ ground 
state and the low-lying $\rm ^2D^o_{5/2,3/2}$ and $\rm ^2P^o_{3/2,1/2}$ metastable states. 
Metastable fractions are inferred from our present work. Resonance energies and quantum defects
 of the prominent Rydberg resonances series identified in the spectra are 
 compared for the $\rm 4p \rightarrow nd$  transitions with the recent ALS experimental 
measurements made on this complex trans-iron element.
\\
\\
\noindent
(Figures are in colour only in the online version)
\end{abstract}
%
%
%
\pacs{32.80.Dz, 32.80.Fb, 33.60-q, 33.60.Fy}

\vspace{0.25cm} 
{\begin{flushleft}
 Short title: Photoionization of Se$^{+}$ ions\\
\vspace{0.25cm}
J. Phys. B: At. Mol. Opt. Phys: \today
\end{flushleft}}
\maketitle
%
%
%
%

%
\section{Introduction}
Recently, detailed photoionization cross-section measurements for the trans-iron element Se$^{+}$,
carried out at the Advanced Light Source (ALS) synchrotron radiation facility
in Berkeley, California were reported in the literature \cite{Esteves2010,sterling11,Esteves11}. 
The photon energy studied in the experiments ranged from 18.0 eV to 31.0 eV 
and was scanned with a nominal spectral resolution of 28 meV. 
Measurements were also made at a higher photon energy resolution of 5.5 meV 
from 17.75 to 21.85 eV spanning the $\rm 4s^24p^3~ ^4S^o_{3/2}$ 
ground-state ionization threshold and the $\rm ^2P^o_3/2$, $\rm ^2P^o_{1/2}$, 
$\rm ^2D^o_{5/2}$,  and $\rm ^2D^o_{3/2}$ metastable state thresholds.
Although detailed resonance structure in the photoionization 
spectra was reported, it was not possible to positively determine the metastable components 
in the ion beam.   Photoionization cross section calculations have been 
made recently on this complex ion using AUTOSTRUCTURE \cite{witthoeft11b} within the 
Multi-Configuration-Breit-Pauli (MCBP) distorted-wave approximation 
for a wider range of energies and for higher charged states. 
  It was concluded from that study on Se ions  \cite{witthoeft11b}
(by  visual comparison) that best agreement with experiment was obtained by weighting the contribution of the
ground configuration states ($\rm ^4S^o_{3/2}$, $\rm ^2D^o_{3/2}$, $\rm ^2D^o_{5/2}$, 
$\rm ^2P^o_{1/2}$, $\rm ^2P^o_{3/2}$) by (0.53, 0.15, 0.05, 0.11, 0.16) respectively. 
As pointed out by Sterling and Witthoeft \cite{witthoeft11b}, R matrix techniques are necessary to reproduce
 resonance interference effects.  In the present work we report on large scale photoionization  
cross section calculations for this complex trans-iron ion 
in both the Breit-Pauli and Dirac Coulomb R-matrix approximations.  
We infer metastable fractions and benchmark our 
theoretical results with the available experimental 
data \cite{Esteves2010,sterling11,Esteves11} providing confidence in the 
data on this trans-iron element for applications \cite{sterling09,witthoeft11a}.  

Photoionization (PI) of atomic ions is an important process in determining the
ionization balance and hence the abundances of elements in
photoionized astrophysical nebulae.  It has recently become
possible to detect neutron(\emph{n})-capture elements (atomic
number $Z>30$, e.g. Se, Kr, Br, Xe, Rb, Ba and Pb) in a large number of ionized nebulae
\cite{sterling07,sharpee07,sterling08}.  These elements are produced by
slow or rapid \emph{n}-capture nucleosynthesis (the
``\emph{s}-process'' and ``\emph{r}-process,'' respectively) \cite{sharpee07,nuceli2001}.
Measuring the abundances of these elements helps to reveal
their dominant production sites in the Universe, as well as
details of stellar structure, mixing and nucleosynthesis
\cite{smith90,cardelli93,wally97,busso99,trav04,herwig05,cowan06,sterling09}. These
astrophysical observations are the motivation to determine the
photoionization and recombination properties of
\emph{n}-capture elements.

Various \emph{n}-capture elements have been detected in the spectra of planetary
nebulae (PNe\footnote{Following astrophysical nomenclature, we abbreviate the  singular
form planetary nebula as ``PN'' and the plural form as ``PNe.''}), the photoionized ejecta of
evolved low- and intermediate-mass stars (1--8 solar masses).  Planetary Nebula progenitor stars may
experience \emph{s}-process nucleosynthesis \cite{busso99,stran06}, in which case their nebulae
will exhibit enhanced abundances of trans-iron elements.  The level of \emph{s}-process
enrichment for individual elements is strongly sensitive to the physical conditions in
the stellar interior \cite{busso99,herwig05}.

The underlying difficulty in studying \emph{s}-process
enrichment in PNe is  due to the large uncertainties (factors of 2--3)
of \emph{n}-capture element abundances derived from the
observational data \cite{sterling11,sterling09}.  There are two main causes for these
uncertainties.  First, the low cosmic abundances
of trans-iron elements, means their emission lines are sufficiently
weak that generally only one or two ions of each element can be
detected in individual PNe.   Second, while
robust ionization corrections can be derived from numerical
simulations of nebulae \cite{ferland98,kallman01}, this method
relies on the availability of accurate atomic data for
processes that affect the ionization equilibrium of each
element.  In photoionized nebulae, these atomic data include
photoionization cross-sections and rate coefficients for
radiative and dielectronic recombination and charge exchange
reactions are unknown for nearly all
\emph{n}-capture element ions.  Uncertainties in the
photoionization and recombination data of \emph{n}-capture
element ions can result in elemental abundance uncertainties of
a factor of two or more \cite{sterling07}.

The present theoretical investigation is a 
study to determine the photoionization and recombination
properties of \emph{n}-capture element ions, motivated by the astrophysical detection of these
species and the importance of their elemental abundances for testing theories of nucleosynthesis and
stellar structure.  Determining the appropriate atomic data 
over the range of energies and temperatures encountered in
astrophysical environs necessitates a predominantly theoretical approach. 
Experimental measurements are helpful to validate theoretical studies,
particularly in the case of complex heavy ionic systems such as low charge states of trans-iron elements.
We note that experimental studies are often limited in the photon energy range studied, 
energy resolution and to the charge states able to be investigated. 
In addition,  many of the experiments carried out are 
contaminated by metastable states and require theoretical 
estimates to help unravel the components in the ion beam.

The prime motivation for choosing Se as the first element for investigation is due to the fact it has been detected in
nearly twice as many PNe (70 in total) as any other trans-iron elements \cite{sterling08}
in addition to recent experimental measurements becoming 
available to compare with \cite{Esteves2010,sterling11}.
Experimental photoionization studies of other astrophysically observed \emph{n}-capture
elements have already been conducted by various groups in the cases of select Kr
\cite{lu06a,lu06b,bizau11} and Xe ions \cite{bizau06,bizau11}.

In this paper we present theoretical determinations of the absolute Se$^+$
photoionization cross-section near the ground-state ionization threshold and compare 
our results with recent experimental measurements \cite{Esteves2010,sterling11} for this trans-iron ion. 
We note that the results from our work can be applied to calibrate a broader
theoretical effort to determine the photoionization and recombination properties of
\emph{n}-capture element ions over a large range of energies and temperatures.  
Our data would be suitable for incorporation within photoionization modelling codes such as Cloudy and 
XSTAR\cite{ferland98, kallman01}, which numerically simulate 
the thermal and ionization structure  of astrophysical nebulae.
Photoionization models can be used to derive accurate
and robust ionization corrections for trans-iron elements, thereby enabling the
accurate determination of their abundances in the Universe. 
We also intend to explore the properties of other members of this iso-electronic sequence, 
for example, Br$^{2+}$ and Rb$^{3+}$ along with other low-charged Se ions in subsequent publications.

The layout of this paper is as follows. Section 2 presents a brief outline of the theoretical work. 
Section 3 presents our theoretical PI cross sections results where comparisons are made 
with the recent Se$^+$ photoionization experimental measurements \cite{Esteves2010,sterling11,Esteves11}. 
Resonance energies and quantum defects for the various  dominant Rydberg series 
(seen in both the experimental and theoretical photoionization spectra) are tabulated and compared. 
Finally in section 4 conclusions are drawn from the present investigation.

\section{Theory}

Relativistic corrections can play an increasingly significant
role for atoms with charge Z~$\ge 18$ \cite{grant07}. It was shown
in recent work on electron-impact excitation of  neutral neon
(Z~${\rm = 10}$)  that a $jK$ coupling
scheme was required to accurately describe even the low lying
energy target levels \cite{ballance04}.   The semi-relativistic Breit-Pauli R-matrix approach 
has been successfully applied repeatably to atomic ions below Z=30 for numerous 
atomic systems by the Iron Project \cite{Hummer:93}.
For atomic ions with $Z>30$, a Dirac-Coulomb approach to electron and photon
interactions is ideal, due to the importance of relativistic and correlation effects
in accurately modelling these systems

 Computationally, the increased number of relativistic orbitals over the corresponding
 non-relativistic orbitals can cause a dramatic increase both in the number
 of integrals required and in the complexity of the angular algebra. 
 We note that in a recent work for the electron-impact excitation 
 of Fe$^{4+}$ \cite{ballance07} and Fe$^{5+}$ \cite{ballance08}
 hundreds of millions of Racah coefficients were calculated for the relativistic treatment
 of this open-d shell system. To address the future challenge of electron scattering in
 which hundreds of levels and thousands of scattering channels are to be considered
 an efficient parallel version \cite{ballance06}  of the DARC
  \cite{norrington87,norrington91,norrington04,darc} suite of codes was developed.
 
\begin{figure}
\includegraphics[width=15cm,height=12cm]{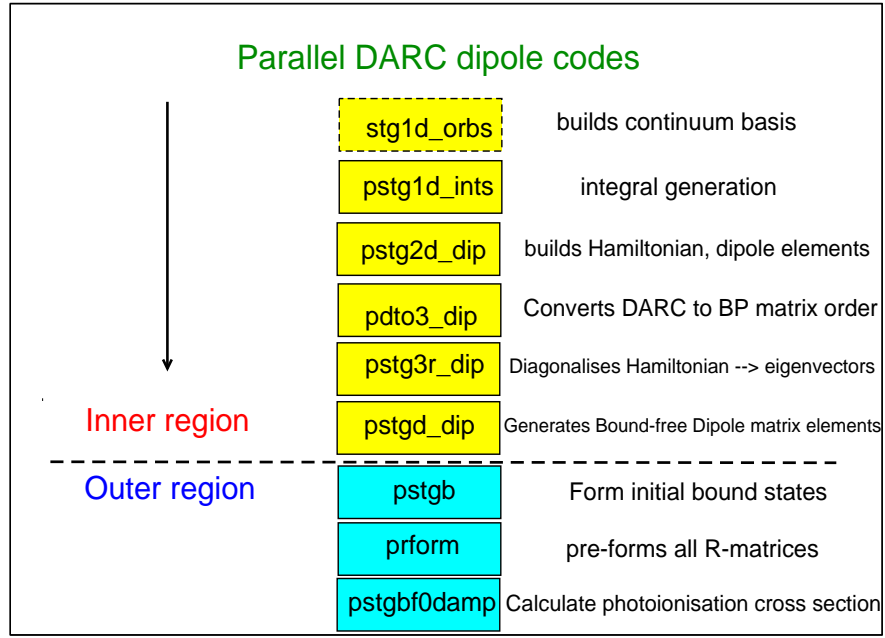}
\caption{Flowchart illustrating the steps necessary
	     to obtain photoionization cross sections with
               the parallel Dirac Atomic R-matrix Codes (DARC).
               The codes are available at http://connorb.freeshell.org/codes/DARCdipole/. 
               \label{DARC}}
\end{figure}

 A natural extension of this work was the formation of the numerous bound-free dipole matrix elements
 required for photoionization in an equally efficient manner, distributed over an equivalent number of processors.
 This enables photoionization and photo-recombination calculations 
 on complex halogen-like ions for example, Ar$^+$, Kr$^+$ and Xe$^+$, 
 to be carried out at the same degree of accuracy \cite{ballance12} as those for electron impact excitation.
 It should be noted that in addition to large scale relativistic photoionization
 calculations, the same dipole matrix elements are the foundation of
 radiationally damped electron-impact excitation studies required in highly charged ions.
 With these extensions to the DARC codes one can address the 
 complex problem of  trans-iron element single photon ionization.   
 The flowchart in figure~\ref{DARC}, with brief annotations beside
 each module of the suite of DARC codes, illustrates the R-matrix sequence 
 required to obtain photoionization cross sections with the parallel 
 {\bf D}irac {\bf A}tomic {\bf R}-matrix {\bf C}odes.
 These recent modifications now include the concurrent formation of every dipole matrix, 
 reducing the total time substantially required for the formation of all dipoles to that required by a single dipole. 
 These developments open the frontier to enabling comprehensive pioneering large scale calculations 
 along iso-nuclear sequences to be performed now within a feasible time frame. 
 
 Using this suite of DARC codes detailed photoionization cross section
  calculations for the Se$^{+}$ ion were performed for 
 the ground and all the excited metastable levels associated with the $\rm 4s^24p^3$ configuration.
 To benchmark our theoretical results we compare them with recent high resolution experimental 
 measurements made at the Advanced Light Source synchrotron 
 radiation facility \cite{Esteves2010,sterling11,Esteves11}.
%
%
    
 In R-matrix theory, all photoionization cross section calculations require 
 the generation of atomic orbitals based primarily on 
 the atomic structure of the residual ion.
 The present theoretical work for the photoionization of the Se$^+$ ion employs
 relativistic atomic orbitals generated for the residual Se$^{2+}$ ion,
 which were calculated using the extended-optimal-level (EOL) procedure within the GRASP
 structure code \cite{dyall89,grant06,grant07}. We explored various
 structure and scattering models of increasing complexity with the R-matrix approach 
 in order to investigate an optimum model for determining the total photoionization cross sections.
 %
%
\begin{figure}
\includegraphics[width=15cm,height=12cm]{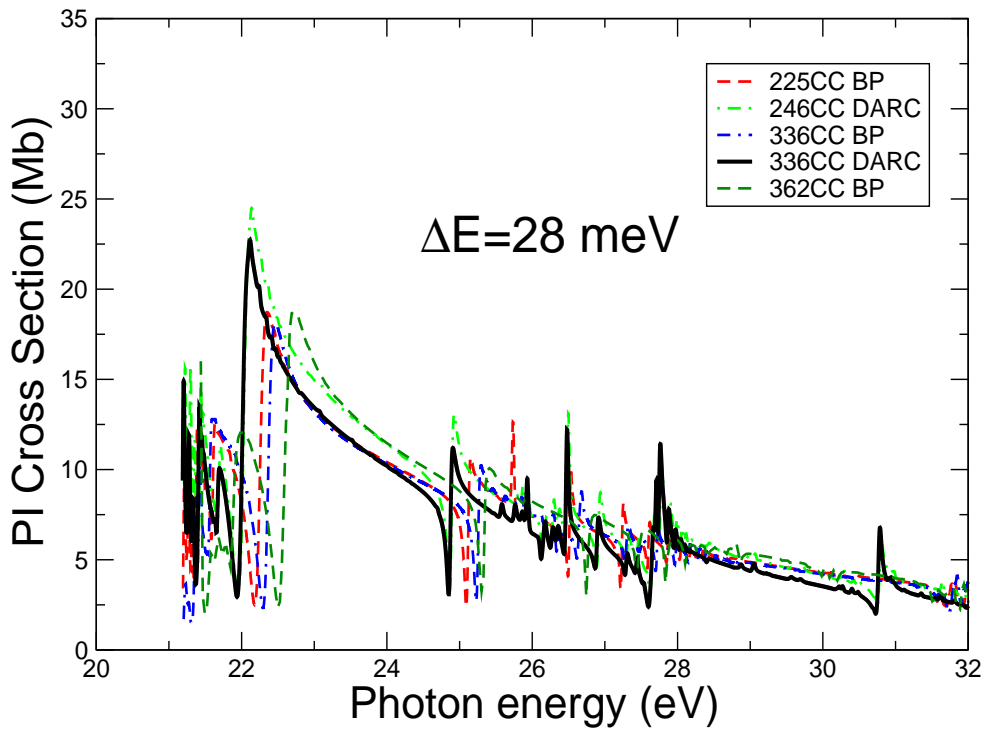}
\caption{(Colour online) Photoionization of the $^4$S$\rm ^o_{3/2}$ ground state of the Se$^+$ ion.   
                The cross section from threshold to a photon energy  of 30 eV  are illustrated.  
               Theoretical results have been obtained from both the DARC and Breit-Pauli
                approximations with an increasing number of levels included in the close-coupling approximations. 
                The theoretical cross sections have been convoluted with a 28 meV FWHM gaussian 
                distribution for comparison purposes. \label{BP_DARC}}
\end{figure}
    
Photoionization cross section calculations for 
 this complex trans-iron system began by including only levels from the $\rm 4s^24p^2$, $\rm 4s4p^3$ 
 and $\rm 4p^4$ configurations which gave rise to 20 $J \pi$ levels. 
 We found this model to be inadequate to represent the complex experimental spectrum 
and it was necessary to extend the theoretical model.
Our first approach was to supplement the configurations of this 20 level model with 
 addition levels from several 3d-hole configurations, $\rm 3d^94s^24p^3$, 
 $\rm 3d^94s4p^4$ and $\rm 3d^94p^5$.  This increased 
 the number of levels in our close coupling expansion to 126  
yielding improved agreement with experiment. 
 Further extensive theoretical models were then
 investigated with the inclusion of levels from configurations involving a $\rm 4d$ orbital
 in our close coupling expansion rather than levels from the 3d hole configurations.
 The first of these models included all 246 levels arising from the eight configurations: 
 $\rm 4s^24p^2$, $\rm 4s4p^3$, $\rm 4p^4$, $\rm 4s^24p4d$, $\rm 4s^24d^2$, $\rm 4p^24d^2$,
 $\rm 4s4p^24d$ and $\rm 4p^34d$ were included in the close-coupling expansion. 
 The second model investigated was when this 246-level model was augmented 
 with levels from an additional $\rm 4s4p4d^2$ configuration,  giving rise to a total of 336 levels.  
 
 The final model augmented this 336-level model with the $\rm 4s^24p5s$, $\rm 4s^24p5p$  and 
 $\rm 4s^24p5d$  configurations which gave rise to a total of 362-levels.  
 Table 1 shows a sample comparison of our results for the lowest 12 levels
 of the residual Se$^{2}$ (Se III)  ion with the available experimental and theoretical work.
 
 Photoionization cross section calculations  were performed in 
 the Dirac-Coulomb and Breit-Pauli approximation with the various models. 
 The advantage of the DARC calculation over the Breit-Pauli 
 result is marginal for the photoionization of outer shell electrons.
 However, it does provide the repeatability of the result using 
 completely independent structure codes and for the most part 
 independent collisional codes to provide the cross section.
 
 Considering the ionization potential between the ground state of Se II 
 and that of Se III is approximately 1.6 Ryds and according to NIST tables
 the $\rm 4s^2 4p 4d$, $\rm4s^2 4p 5\ell$ levels are over 1 Ryd above the Se III ground state. 
 This gives a photon energy of greater than 36 eV to photoexcite to 
 either the $\rm 4d^24p4d$ or the $\rm 4s^24p5s$ levels. Therefore, 
 their influence on the ground state photoionization cross section 
 within the energy range presented in this paper, is solely 
in terms of changes in structure through CI mixing.

 Figure 2 shows that the inclusion of the additional 5s, 5p and 5d orbitals in the basis set 
 has a minimal effect on the total photoionization cross section for the $\rm ^4S^o_{3/2}$ 
 ground state over the energy range investigated and for energies less than 31 eV.
 However, these comparisons with experiment are slightly misleading 
 as the energy levels of the DARC R-matrix calculation are shifted to experimental 
 values prior to the matrix diagonalization. 
 Since converged results are obtained within the 336  level model we used this approximation 
 to determine all the remaining  PI cross sections for the metastable states with the DARC codes.

  \begin{table}
\caption{Comparison of the theoretical energies using GRASP and the Breit-Pauli approximation 
                from the 362-level model with previous work,  relative energies are in Rydbergs.
	       A sample of the first 12 levels of the Se III ion are compared with AUTOSTRUCTURE 
	        \cite{witthoeft11b} and the NIST \cite{Ralchenko2010} tabulated values.}
\begin{tabular}{cccccccc}
\br
Level       &  Config. &  Term$^{\rm a}$            & Energy  		& Energy 		&Energy                             & Energy                & $\Delta (\%)^{\rm b}$ \\
		&			&			     &  NIST			& GRASP 		& Breit-Pauli$^{\rm d}$    & AUTOS$^{\rm e}$				 & \\	
\mr
 1  & 4s$\rm ^2$4p$\rm ^2$ 		&  $\rm ^3$P$\rm _0$          & 0.000000 		&0.000000	&0.00000		& 0.00000 	                    &     0.0 \\  
 2  & 4s$\rm ^2$4p$\rm ^2$ 		&  $\rm ^3$P$\rm _1$          & 0.015870		&0.015061	&0.01419		& 0.01580	 			  &    5.1 \\
 3  & 4s$\rm ^2$4p$\rm ^2$ 		&  $\rm ^3$P$\rm _2$          & 0.035880 		&0.035409	&0.03347		& 0.03620				  &    1.3 \\ 
 4  & 4s$\rm ^2$4p$\rm ^2$		&  $\rm ^1$D$\rm _2$          & 0.118760	 	&0.140979	&0.13835		& 0.14370 			  &  18.7 \\
 5  & 4s$\rm ^2$4p$\rm ^2$ 		&  $\rm ^1$S$\rm _0$          & 0.259070 		&0.285184	&0.27116		& 0.25950	 			  &  10.1 \\
 6 & 4s$\rm ^1$4p$\rm ^3$		&  $\rm ^5$S$\rm ^o_2$     &\, 0.579910$^c$	&0.566575	&0.55304		& --		 	           	  &    2.3  \\
 7  & 4s$\rm ^1$4p$\rm ^3$ 		&  $\rm ^3$D$\rm ^o_1$     & 0.830080		&0.853381	&0.81498		& 0.83080 		  	  &    2.8 \\
 8  & 4s$\rm ^1$4p$\rm ^3$ 		&  $\rm ^3$D$\rm ^o_2$     & 0.844950 		&0.854258	&0.81543		& 0.84500 			  &    1.1 \\
 9  & 4s$\rm ^1$4p$\rm ^3$ 		&  $\rm ^3$D$\rm ^o_3$     & 0.879810 		&0.859092	&0.81900		& 0.87980 			  &    2.3 \\
10 & 4s$\rm ^1$4p$\rm ^3$ 		&  $\rm ^3$P$\rm ^o_0$     & 0.970335 		&0.991048	&0.94428		& 0.96750				  &    2.1 \\
11 & 4s$\rm ^1$4p$\rm ^3$ 		&  $\rm ^3$P$\rm ^o_2$     & 0.970636 		&0.994054	&0.94614		& 0.96540 			  &    2.4 \\
12 & 4s$\rm ^1$4p$\rm ^3$     	         &  $\rm ^3$P$\rm ^o_1$     & 0.971329 		&0.992406	&0.94616		& 0.96760 			  &    2.2\\
\mr
\end{tabular}
$^{\rm a}$ $^{2S+1}L^{\pi}_J$\\
$^{\rm b}$ GRASP,  absolute percentage difference relative to NIST values\\
$^{\rm c}$ Esteves PhD thesis \cite{Esteves2010}\\
$^{\rm d}$ Breit-Pauli  \\
$^{\rm e}$ Sterling and Witthoeft  \cite{witthoeft11b}\\
\end{table}

 The R-matrix boundary radius of 8 Bohr radii  was sufficient to envelop
 the radial extent of all the atomic orbitals of the residual Se$^{2+}$ ion. A basis of 16 continuum
 orbitals was sufficient to span the incident experimental photon energy
 range of up to 32 eV. The 336-state model produced a maximum of
 1758 coupled channels in our scattering work with Hamiltonian matrices  of dimension
of the order of 28,300 by 28,300 in size. Due to dipole selection rules, 
 for total ground state photoionization we need only consider the three
 bound-free dipole matrices, $\rm 2J^{\pi}=3^o \rightarrow 2J^{\pi}=1^e,3^e,5^e$ whereas 
 for the excited metastable states one requires $\rm 2J^{\pi}=5^o \rightarrow 2J^{\pi}=3^e,5^e,7^e$ the
 $\rm 2J^{\pi}=3^o \rightarrow 2J^{\pi}=1^e,3^e,5^e$ and $\rm 2J^{\pi}=1^o \rightarrow 2J^{\pi}=1^e,3^e$ bound-free dipole matrices.
 For each initial state the outer region electron-ion collision 
problem was solved (in the resonance region below and
 between all thresholds) using a suitably chosen fine
energy mesh of 5$\times$10$^{-7}$ Rydbergs ($\approx$ 6.8 $\mu$eV) 
to fully resolve all the extremely fine resonance
structure in the appropriate photoionization cross sections. 
The multi-channel R-matrix eigenphase derivative (QB) 
technique (applicable to atomic and molecular complexes) 
of Berrington and co-workers \cite{keith1996,keith1998,keith1999} 
was used to determine the resonance parameters. In the QB method, the resonance width $\Gamma$
is determined from the inverse of the energy derivative of the eigenphase sum $\delta$ 
at the position of the resonance energy $\epsilon_r$ via
\begin{equation}
\Gamma = 2\left[{\frac{d\delta}{dE}}\right]^{-1}_{E=\epsilon_r} = 2 [\delta^{\prime}]^{-1}_{E=\epsilon_r} \quad.
\end{equation}
Prominent photoexcitation-autoionization Rydberg resonances 
series are analyzed and identified as $\rm 4p \rightarrow nd$  transitions, and 
the metastable fractions of the experimental Se$^+$ ion beam are estimated.
Finally, in order to compare directly with recent experimental
 work on this system \cite{Esteves2010,sterling11,Esteves11}, 
the theoretical cross sections were convoluted with a gaussian
function of appropriate width  (FWHM of 28 meV and 5.5 meV respectively)  and weighted accordingly 
in order to simulate the energy resolution of the measurements 
performed at the Advanced Light Source (ALS) synchrotron radiation facility.
%
%
\begin{figure}
\includegraphics[width=15cm,height=12cm]{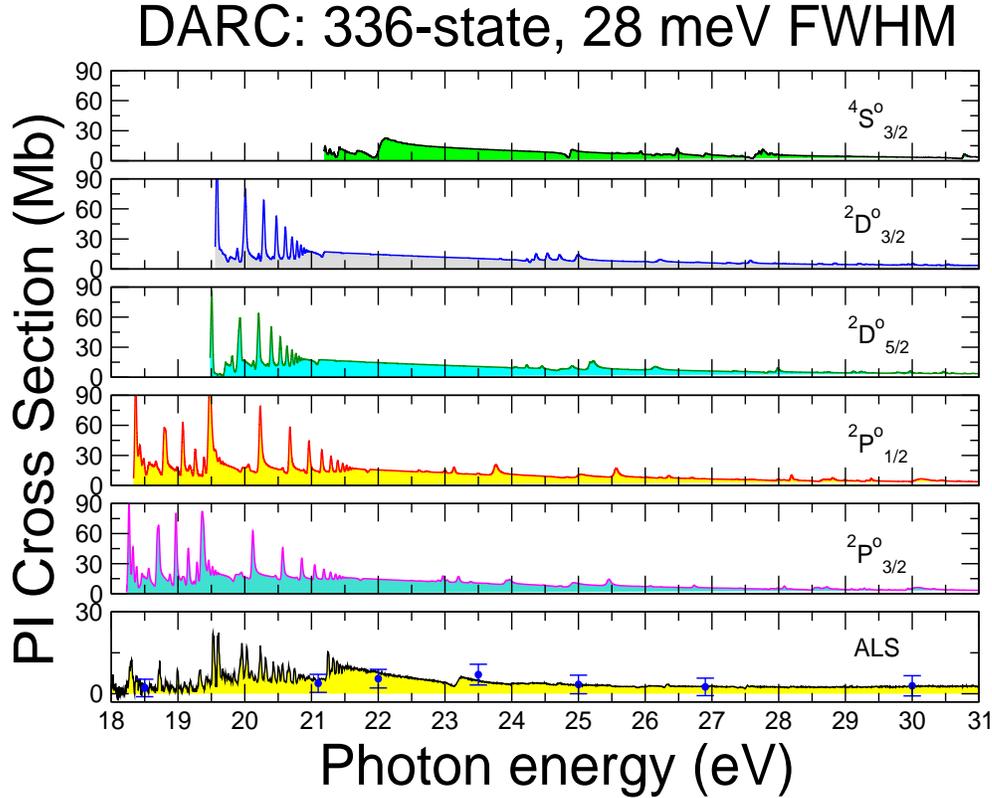}
\caption{(Colour online) Theoretical photoionization cross sections 
                (336-level close-coupling DARC calculations) for the $\rm ^4S^o_{3/2}$ 
                ground state,  $\rm ^2D^o_{3/2}$,  $\rm ^2D^o_{5/2}$,
                $\rm ^2P^o_{3/2}$ and $\rm ^2P^o_{1/2}$ metastable states 
                arising from the $\rm 4s^24p^3$ configuration
               are illustrated (Top 5 panels). The theoretical results were 
               convoluted with a gaussian of  28 meV  FWHM to simulate the photon energy
               resolution of the experimental measurements (bottom panel) 
	     taken at the Advanced Light Source synchrotron radiation facility. \label{StackThrExp}}
\end{figure}

%
%
\begin{figure}
\includegraphics[width=15cm,height=12cm]{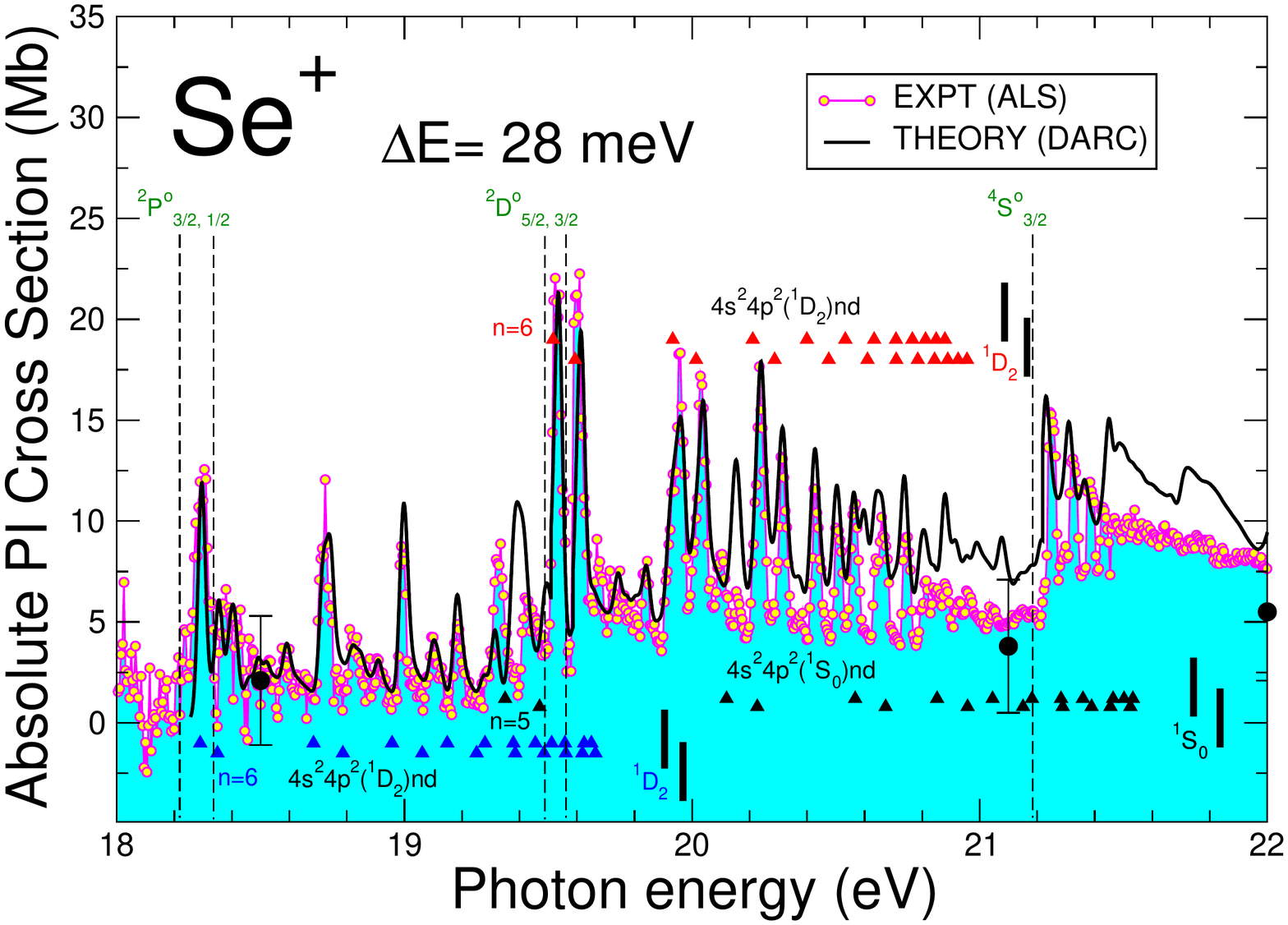}
\caption{(Colour online) Comparison of ALS photoionization cross-section measurements
                from 18 eV to 21 eV taken at 28 meV (shaded area)
                with the 336-state close-coupling DARC   calculations. 
		The theoretical cross sections (solid black line) for each of the five
                initial states have been convoluted with a 28 meV FWHM gaussian
		and a weighting of the states (see text for details) to 
		simulate the measurements.\label{theo_meta_expt} 
		The absolute measurements  (solid black circles) 
                   have been obtained with a larger energy step. 
	          The error bars give the total uncertainty of the experimental data.
	          The various Rydberg series are indicated }
\end{figure}

%
%
\begin{figure}
\includegraphics[width=15cm,height=12cm]{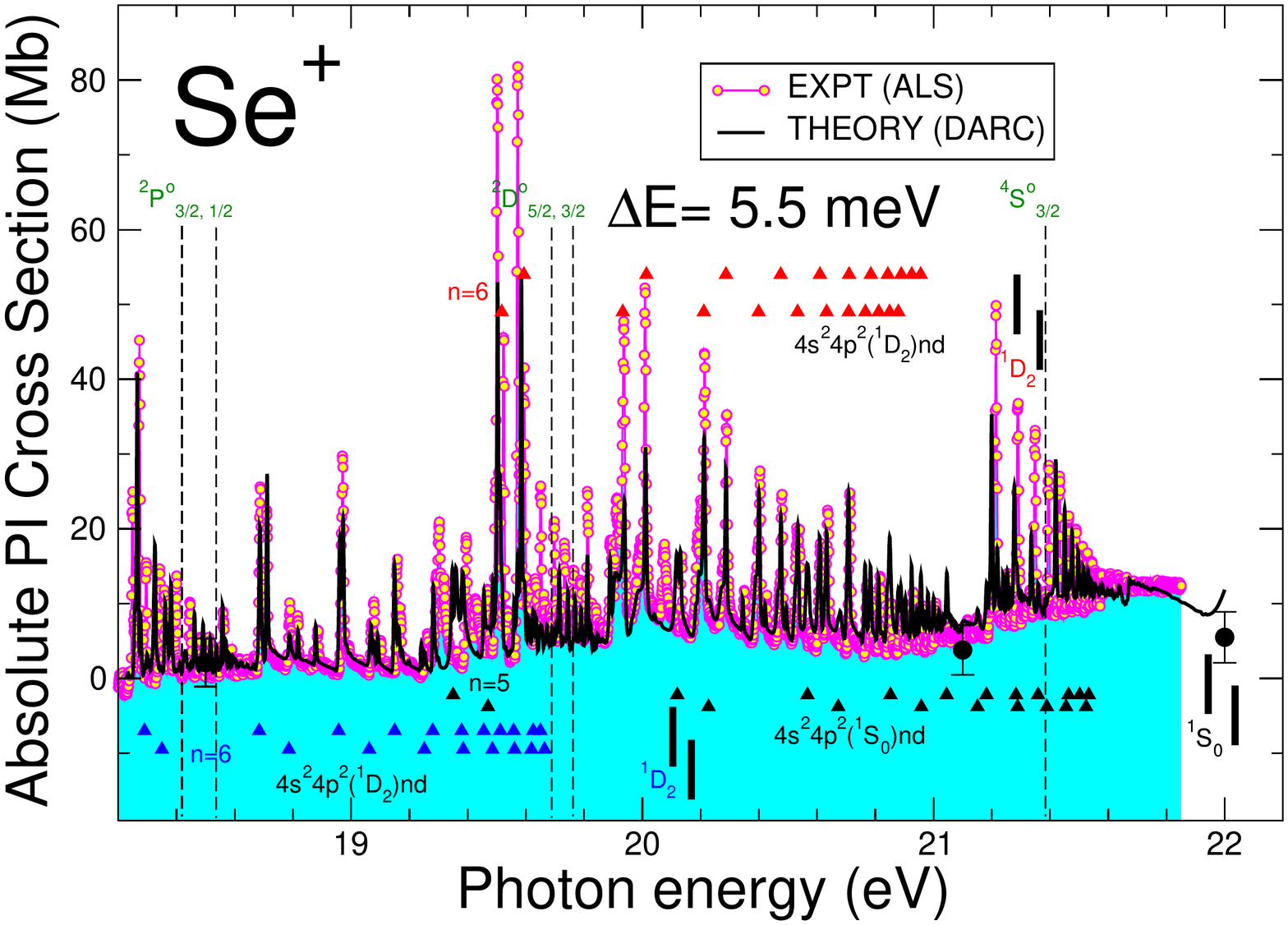}
\caption{(Colour online) Comparison of ALS photoionization cross-section measurements
                from 18 eV to 21.8 eV taken at 5.5 meV (shaded area)
                with the 336-state close-coupling DARC   calculations. 
		The theoretical cross sections (solid black line) for each of the five
                initial states have been convoluted with a 5.5 meV FWHM gaussian
		and a weighting of the states (see text for details) to 
		simulate the measurements.\label{highres} 
		The absolute measurements  (solid black circles) 
                   have been obtained with a larger energy step. 
	          The error bars give the total uncertainty of the experimental data. 
	          The various Rydberg series are indicated. }
\end{figure}
		
%
%
\begin{figure}
\includegraphics[width=15cm,height=12cm]{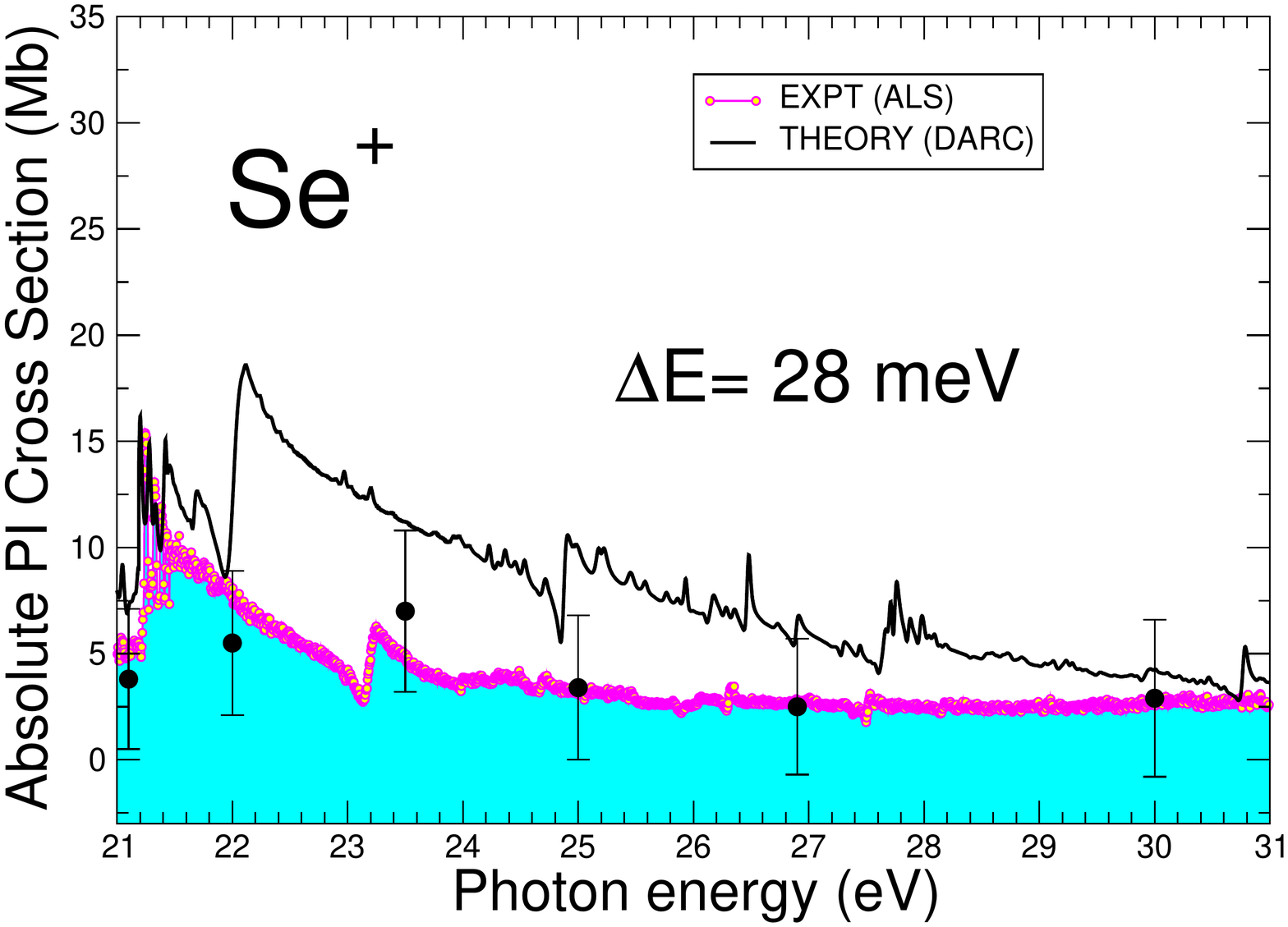}
\caption{(Colour online) Comparison of ALS photoionization cross-section measurements
                from 21 eV to 31 eV taken at 28 meV (shaded area)
                with the 336-state close-coupling DARC calculations. 
		The theoretical cross sections (solid black line) for each of the five
                initial states have been convoluted with a 28 meV FWHM gaussian
		and a weighting of the states (see text for details) to 
		simulate the measurements.\label{theo_expt}
		The absolute measurements  (solid black circles) 
                   have been obtained with a larger energy step. 
	          The error bars give the total uncertainty of the experimental data.}
\end{figure}

%
%

\begin{figure}
\includegraphics[width=15cm,height=12cm]{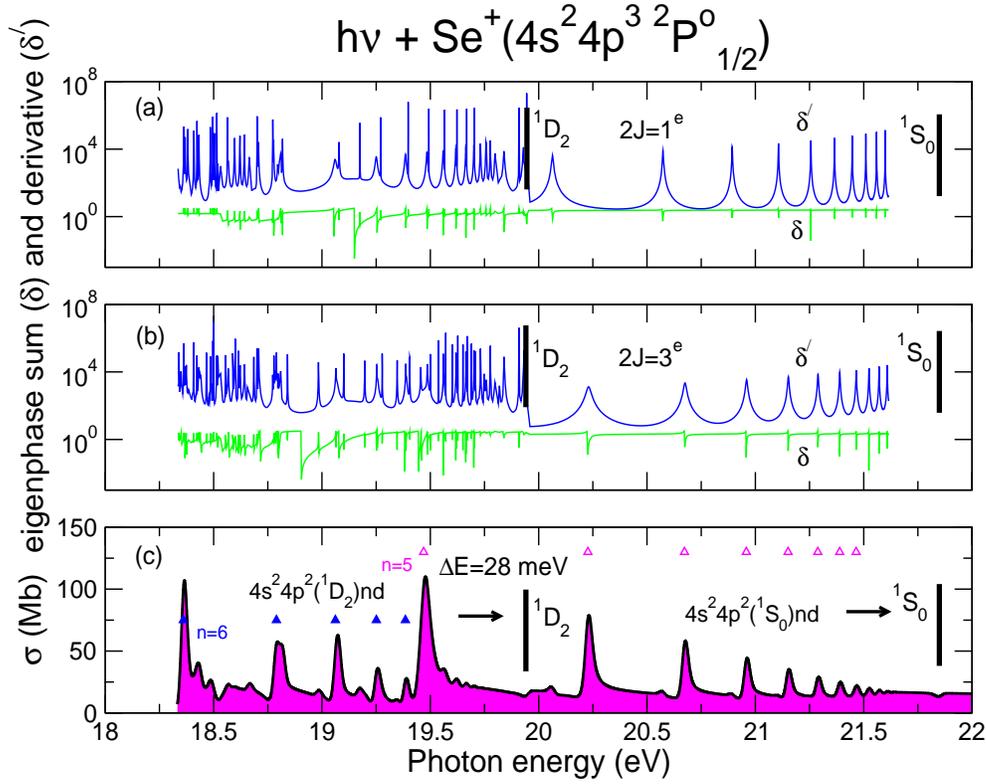}
\caption{(Colour online) Eigenphase sum $\delta$ and its derivative $\delta^{\prime}$ 
                (for each 2J$^{\pi}$ symmetry contributing to the PI cross section $\sigma$ for the $\rm 4s^24p^3~^2P^o_{1/2}$
                 initial state) as a function of photon energy in the region below 
                 the $\rm ^1D_2$ and $\rm ^1S_0$ excited state thresholds of the residual Se$^{2+}$ ion for
                (a) 2J=1$\rm ^{e}$ symmetry,  (b) 2J=3$\rm ^{e}$ symmetry  
                 and (c) PI cross section $\sigma$ for the $\rm ^2P^o_{1/2}$ metastable state 
                convoluted with a gaussian of 28 meV FWHM (solid black line).
                The prominent Rydberg resonances series in the 
                PI cross section have been identified
               (open and solid triangles) as members of two different Rydberg
              autoionizing  series 
              converging to the $\rm ^1D_2$ and $\rm ^1S_0$ excited
               states of the residual Se$^{2+}$ ion indicated by thick vertical lines.\label{fig7}}
  \end{figure}

%
%

\begin{figure}
\includegraphics[width=15cm,height=12cm]{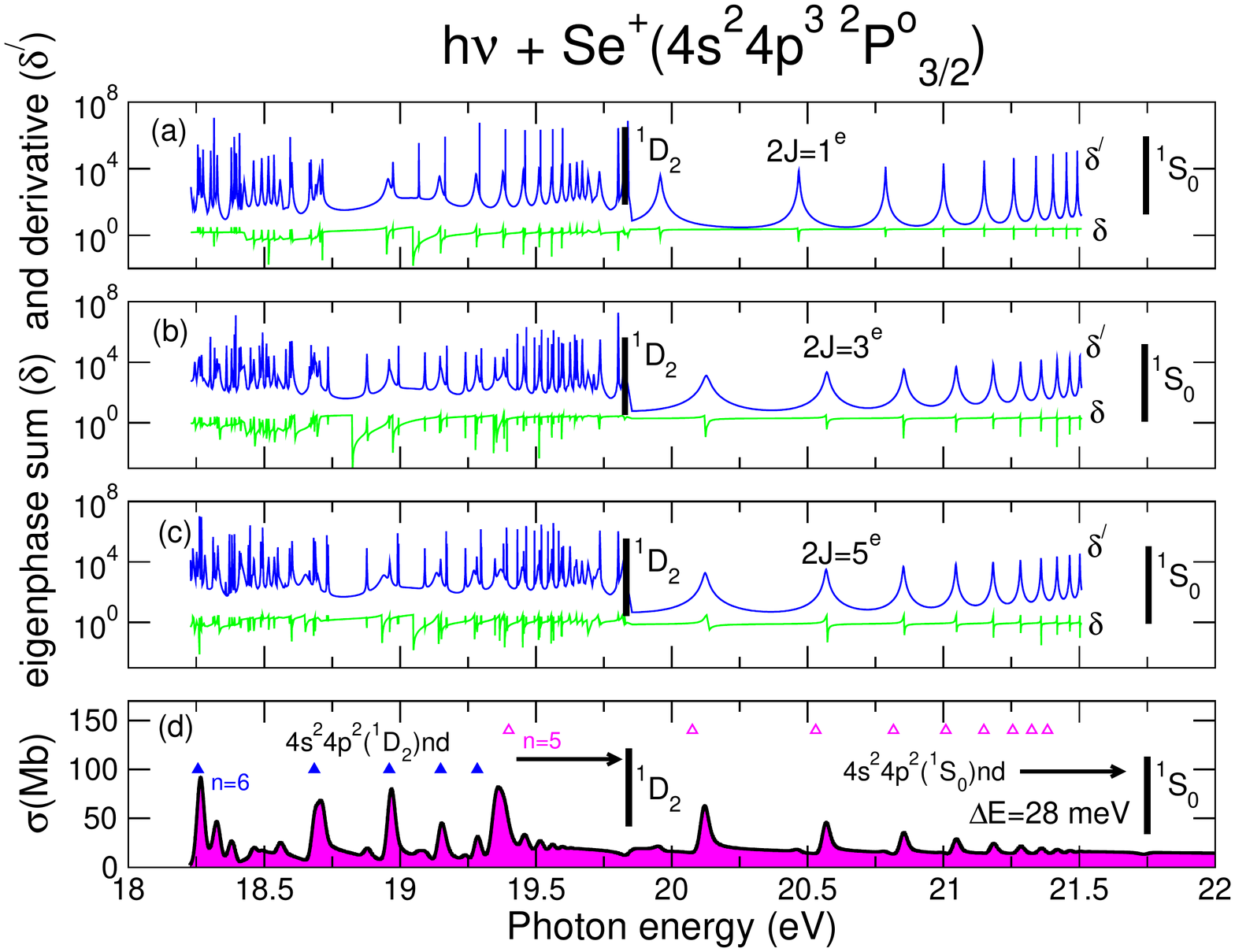}
\caption{(Colour online) Eigenphase sum $\delta$ and its derivative $\delta^{\prime}$ 
                (for each 2J$^{\pi}$ symmetry contributing to the PI cross section $\sigma$ for the $\rm 4s^24p^3~^2P^o_{3/2}$
                 initial state) as a function of photon energy in the region below
                  the $\rm ^1D_2$ and $\rm ^1S_0$ excited state thresholds of the residual Se$^{2+}$ ion for
                (a) 2J=1$\rm ^{e}$ symmetry,  (b) 2J=3$\rm ^{e}$ symmetry, 
                (c) 2J=5$\rm ^{e}$ symmetry and (d) PI cross section $\sigma$
                for the $\rm ^2P^o_{3/2}$ metastable state  
                convoluted with a gaussian of 28 meV FWHM (solid black line).
                The prominent Rydberg resonances series in the 
                PI cross section have been identified
               (open and solid triangles) as members of two different Rydberg
              autoionizing series 
              converging to the $\rm ^1D_2$ and $\rm ^1S_0$ excited
               states of residual Se$^{2+}$ ion indicated by thick vertical lines.\label{fig8}}
  \end{figure}

%
%

\begin{figure}
\includegraphics[width=15cm,height=12cm]{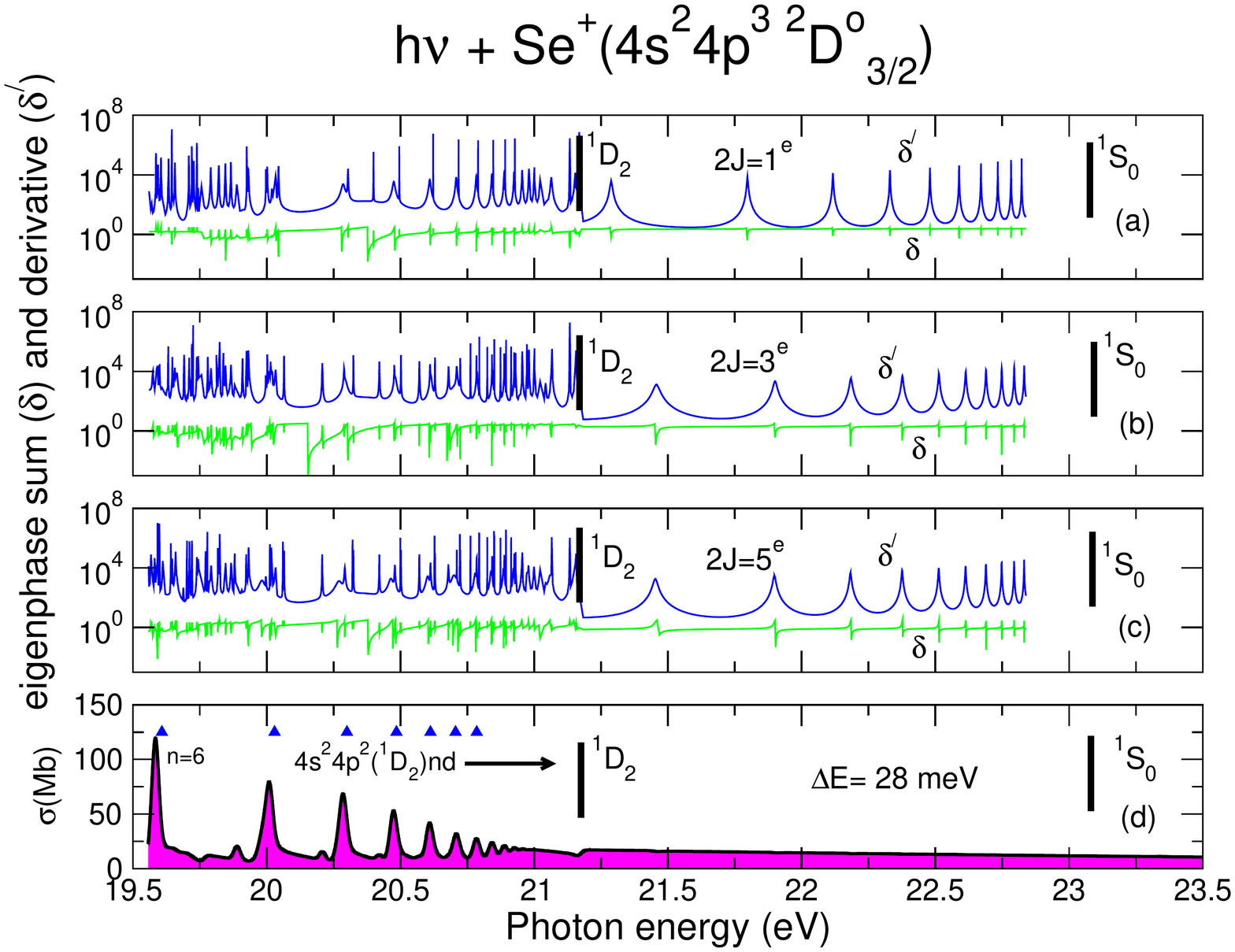}
\caption{(Colour online) Eigenphase sum $\delta$ and its derivative $\delta^{\prime}$ 
                (for each 2J$^{\pi}$ symmetry contributing to the PI cross section $\sigma$ for the $\rm 4s^24p^3~^2D^o_{3/2}$
                 initial state)  as a function of photon energy in the region below
                  the $\rm ^1D_2$ and $\rm ^1S_0$ excited state thresholds the residual Se$^{2+}$ ion for
                (a) 2J=1$\rm ^{e}$ symmetry,  (b) 2J=3$\rm ^{e}$ symmetry, 
                (c) 2J=5$\rm ^{e}$ symmetry and (d) PI cross section $\sigma$
                for the $\rm ^2D^o_{3/2}$ metastable state  
                convoluted with a gaussian of 28 meV FWHM (solid black line).
                The prominent Rydberg resonances series in the 
                PI cross section have been identified
               (solid triangles) as members of a Rydberg
              autoionizing series 
              converging to the $\rm ^1D_2$ threshold of the residual Se$^{2+}$ ion.  
              The weak resonance series converging to the 
              $\rm ^1S_0$  threshold is not observed in the PI cross sections. \label{fig9}}
  \end{figure}

%
%

\begin{figure}
\includegraphics[width=15cm,height=12cm]{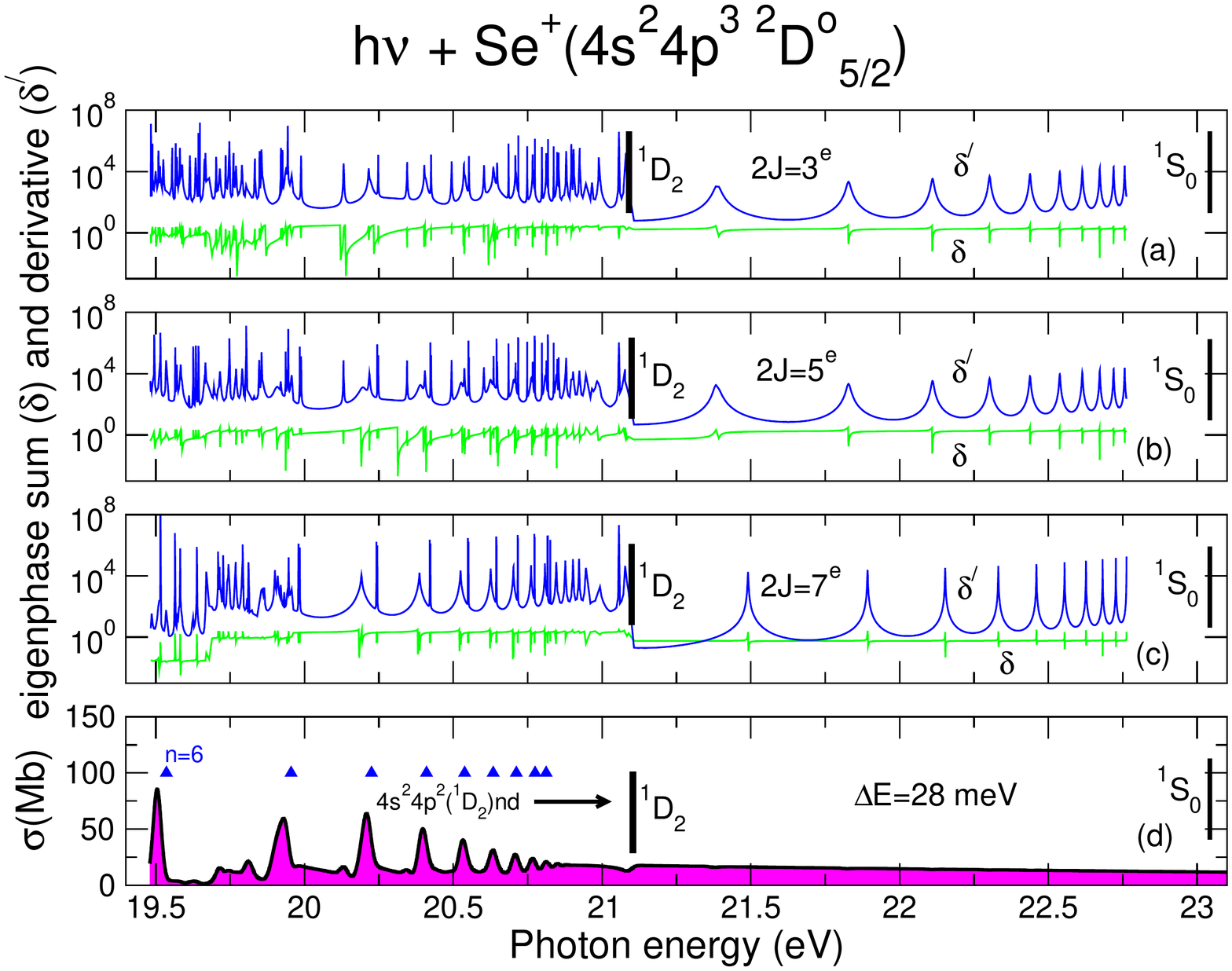}
\caption{(Colour online) Eigenphase sum $\delta$ and its derivative $\delta^{\prime}$ 
                (for each 2J$^{\pi}$ symmetry contributing to the PI cross section $\sigma$ for the $\rm 4s^24p^3~^2D^o_{5/2}$
                 initial state) as a function of photon energy in the region below
                  the $\rm ^1D_2$ and $\rm ^1S_0$ excited state thresholds of the residual Se$^{2+}$ ion for
                (a) 2J=3$\rm ^{e}$ symmetry,  (b) 2J=5$\rm ^{e}$ symmetry, 
                (c) 2J=7$\rm ^{e}$ symmetry and (d) PI cross section $\sigma$
                for the $\rm ^2D^o_{5/2}$ metastable state  
                convoluted with a gaussian of 28 meV FWHM (solid black line).
                The prominent Rydberg resonances series in the 
                PI cross section have been identified
               (solid triangles) as members of a Rydberg
              autoionizing series 
              converging to the $\rm ^1D_2$.  The weak resonance series converging to the 
              $\rm ^1S_0$  threshold is not observed in the PI cross sections. \label{fig10}}
  \end{figure}

%
%

\begin{figure}
\includegraphics[width=15cm, height=12cm]{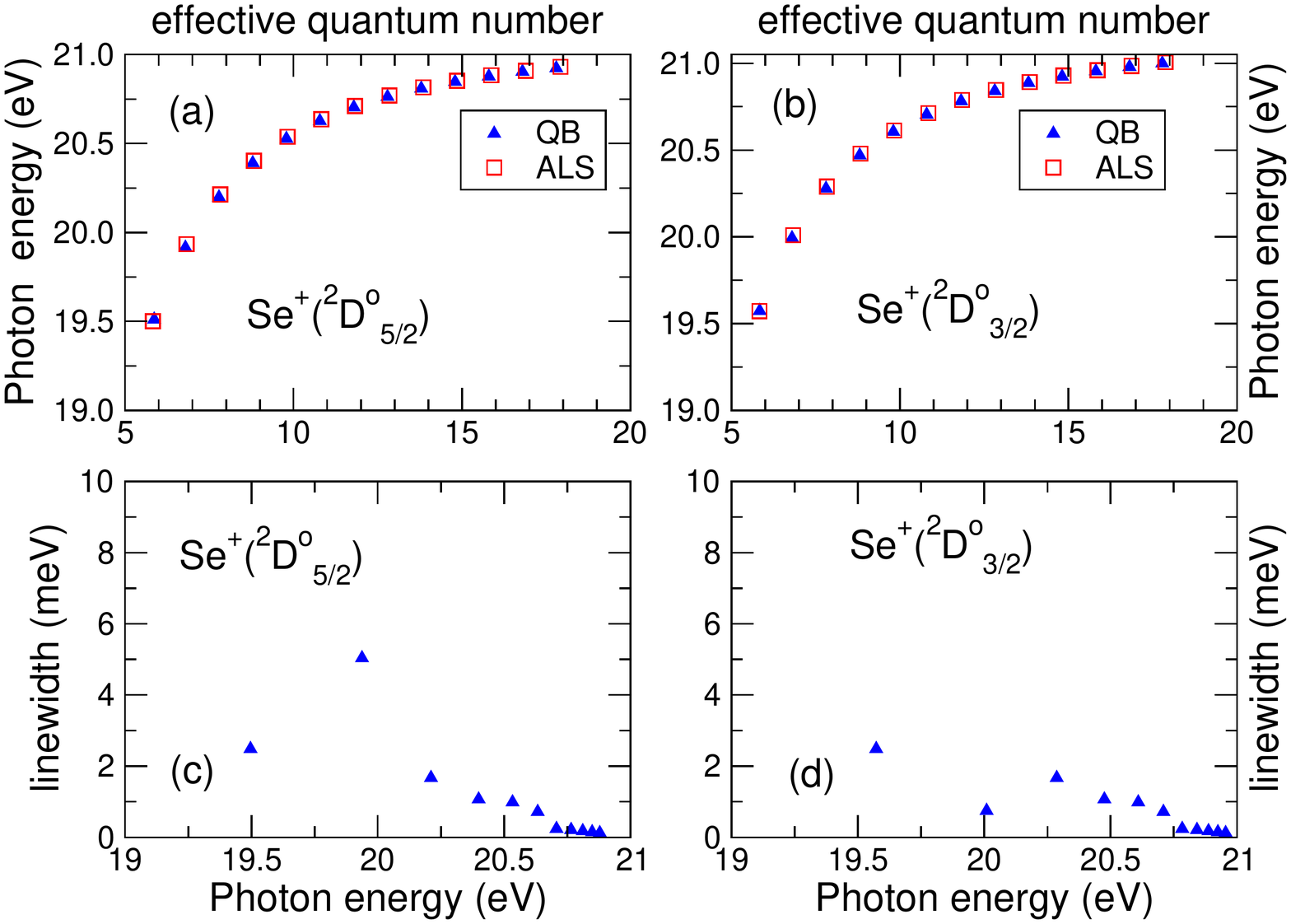}
\caption{(Colour online) Effective quantum number $\nu$ = n - $\mu$ and the 
				       linewidth $\Gamma$ (meV) versus photon energy (eV) 
               				for the  $\rm \:4s^24p^2(^1D_2)nd$ resonances lying  
					below the  $\rm ^1D_2$ threshold of the residual Se$^{2+}$ ion.
             				 Experimental data (open red squares, ALS \cite{Esteves2010,sterling11}) 
              				and theoretical estimates (solid blue triangles, QB);
               				(a)  effective quantum numbers ($\nu$) for the $\rm \:4s^24p^2(^1D_2)nd$ series 
					originating from the initial Se$^+$($\rm ^2D^{o}_{5/2})$ state; 
               				(b)  effective quantum numbers ($\nu$) for the $\rm \:4s^24p^2(^1D_2)nd$ series 
					originating from the initial Se$^+$($\rm ^2D^{o}_{3/2})$ state;
               				(c) $\rm \:4s^24p^2(^1D_2)nd$,  resonance  linewidths ($\Gamma$) 
                    			 originating from the initial Se$^+$($\rm ^2D^{o}_{5/2})$ state and 
               				(d) $\rm \:4s^24p^2(^1D_2)nd$, resonance linewidths ($\Gamma$) 
                   			 originating from the initial Se$^+$($\rm ^2D^{o}_{3/2})$ state. \label{fig11}}
\end{figure}

\section{Results and discussion}
Figure \ref{BP_DARC} shows the convergence of the PI results  
obtained from the Breit-Pauli and Dirac-Coulomb (DARC) approximations 
for the $\rm 4s^24p^3$\,$\rm\,^4S^o_{3/2}$ ground state using an n=4 basis.
 The theoretical PI cross section results are illustrated  for the $\rm 4s^24p^3$\,$\rm\,^4S^o_{3/2}$ 
 ground state of this complex trans-iron element 
 with an increasing number of levels retained  in the close-coupling expansion. 
 As  illustrated in figure \ref{BP_DARC}, the 336-state model yields a suitable 
 description of these half collision processes for the Se$^+$ ion, where
 converged results are obtained with a larger close-coupling approximation.
At this level of approximation it ensures consistency in our theoretical work. 
Using this 336-level model we obtained the PI cross sections for all the remaining metastable levels 
associated with the $\rm 4s^24p^3$ configuration within the Dirac-Coulomb (DARC) approximation.
The theoretical PI cross section results for the individual levels arising from the $\rm 4s^24p^3$ configuration 
 over the photon energy range 18 eV to 31 eV from the Dirac-Coulomb (DARC) approximation
 are shown in figure \ref{StackThrExp} where the results are convoluted with a gaussian distribution of 28 meV FWHM 
 in order to compare with the high resolution ALS experimental results.
The ALS measurements for Se$^{+}$ \cite{sterling11} were taken over the 18 eV to
31 eV photon energy range at an energy resolution of 28 meV,
these measurements are illustrated in the bottom panel of figure~\ref{StackThrExp}. 
The absolute photoionization cross sections were measured at discrete photon energies and are
shown in the figure by solid circles with their associated uncertainties. 

To determine the fractions of metastable ions 
present in the experimental ion  beam photoionization cross section calculations were performed
 using the Dirac-Coulomb R-matrix method for both 
 the $\rm ^4$S$\rm ^o_{3/2}$ ground state and the excited metastable states 
 ( $\rm ^2$D$\rm ^o_{3/2}$, $\rm ^2$D$\rm ^o_{5/2}$,   $^2$P$\rm ^o_{1/2}$ and $\rm ^2$P$\rm ^o_{3/2}$ ) 
associated with the  4s$^2$4p$^3$  configuration.
The individual fractions were estimated by first scaling the highest-lying
metastable state to the absolute experimental cross section.
Each subsequent state was then scaled such that the sum of the
theoretical cross sections matched the experiment as closely as
possible, with all four metastable states together with the ground state
fractions ultimately constrained to sum to unity.  
Using this technique the various fractions were determined and it was  
found that a non-statistical distribution of the ground and metastable states gave
best agreement with experiment by weighting the contribution of the
$\rm ^4$S$\rm ^o_{3/2}$,  $\rm ^2$D$\rm ^o_{3/2}$, 
$\rm ^2$D$\rm ^o_{5/2}$,   $^2$P$\rm ^o_{1/2}$ 
and $\rm ^2$P$\rm ^o_{3/2}$  states 
by (0.52, 0.13, 0.19, 0.03 and 0.13) respectively. 

As noted previously, photoionization cross section calculations have been 
made recently on this complex ion 
using AUTOSTRUCTURE \cite{witthoeft11b} within the 
Multi-Configuration-Breit-Pauli (MCBP) distorted-wave approximation 
for a wider range of energies and for higher charged states of the ion. 
Using this distorted wave approximation the authors
find that the direct PI cross sections for these Se ions differ
by 30\% to 50\% when compared to the available ALS experimental measurements. 
  It was concluded from that study on Se ions  \cite{witthoeft11b}
(by  visual comparison) that best agreement with experiment was obtained by weighting the contribution of the
ground configuration states ($\rm ^4S^o_{3/2}$, $\rm ^2D^o_{3/2}$, $\rm ^2D^o_{5/2}$, 
$\rm ^2P^o_{1/2}$, $\rm ^2P^o_{3/2}$) by (0.53, 0.15, 0.05, 0.11, 0.16) respectively. 
Therefore a non statistical distribution of the PI cross sections in both theoretical approaches 
is seen to provide the best agreement with experiment.

Furthermore the Fano profile in the PI cross section near  23.0 eV is not reproduced
 by the AUTOSTRUCTURE calculations \cite{witthoeft11b} 
 (R matrix techniques are necessary to reproduce
 resonance interference effects), and the lack of this
 interference causes the discrepancy in the direct cross section
 in this energy region. The calculated and experimental Se$^{+}$ direct
 cross sections agree to within 30\% to 50\% over the range of
 experimental energies, except above 30 eV, where the experimental
 measurements are more uncertain since only a single
 absolute measurement was made above 27 eV (compared to
 absolute measurements at lower energies, where independent estimates
 were performed 2 to 3 times for each energy). 
 Therefore, they report uncertainties in the experiment of over 50 \% 
 in the high energy region. 
As R matrix techniques are necessary to reproduce
resonance interference effects  we have carried out 
these calculations in the present investigation.
 
In figure~\ref{theo_meta_expt},  the rich resonance
structure observed in the experimental spectra is generally well represented in the metastable
region by theory. This is still apparent when the theoretical work is compared with
the experimental measurements taken at an even higher resolution \cite{Esteves11}. 
Figure~\ref{highres}  shows the ALS experimental measurements taken at 
a higher resolution of 5.5 meV (from 18 eV to 21.8 eV)  together with the theoretical results 
convoluted with a gaussian of the same FWHM  to simulate 
the experimental resolution and non-statistically weighted as before. 
Here again there is good agreement 
between the experimental and theoretical results.    

In the energy region 21 eV to 31 eV,  (figure~\ref{theo_expt}) the theoretical results are larger 
than experiment. The solid line corresponds to the sum of the ground and metastable state
fractions of the theoretical calculations. 
As indicated in the  ALS experimental studies for Se$^{+}$  \cite{Esteves2010,sterling11},
contamination from higher order radiation produced by the undulator beamline 
may be a possible cause for uncertainties. 
Previous experiments on Xe$^{3+}$ \cite{Emmons05} 
estimated a lower fraction of 2 \% at around 40 eV. 
The measurements  on Se$^{+}$   \cite{Esteves2010,sterling11} are between 
these two photon energies and it was concluded in that work \cite{Esteves2010,sterling11}
the maximum contamination is no greater than 4\%. 
At lower photon energies, the contamination of higher
order radiation is expected to be larger, but not by more than
a factor of 2Ð3 compared to the contamination at 40 eV. 
It was concluded from the Se$^+$ ALS measurements  \cite{Esteves2010,sterling11} that the experimental 
uncertainties on the absolute measurements made are 
estimated to be 30\% including the possible contamination 
of the photon beam by higher order radiation.

Earlier studies on the photoionization spectrum of O$^+$ at the Advanced Light Source 
synchrotron radiation facility \cite{aguilarO+:apjs:03}  were used to identify the
 resonance structure due to the metastable states \cite{Esteves2010,sterling11}. 
This is apparent when the spectra of O$^+$ and Se$^+$ are compared, 
the resemblance is  remarkable but not totally unexpected. 
These respective ions have an  $\rm np^3$ configuration (n=2 for O$^+$ and  n=4 for Se$^+$) 
in their outermost shell giving rise in each case to the same term designation  for
the $\rm ^4S^{o}_{3/2}$ ground state  and the $\rm ^2D^{o}_{3/2, 5/2}$ and $\rm ^2P^o_{1/2, 3/2}$  
metastable states.  In the case of O$^+$, the region of the spectrum covering the energy
range from the $\rm ^2P^o_{1/2, 3/2}$ threshold to the $\rm ^4S^o_{3/2}$ threshold
consisted of very strong resonances due to 2$\rm p$ $\rightarrow$
$\rm nd$ electron excitations of the metastable ions as well as
weak resonances due to 2$\rm p$ $\rightarrow$ $\rm ns$ transitions. 
Similarly,  the Se$^+$ spectra, shows the corresponding 4$\rm p$ $\rightarrow$ $\rm nd$ 
transitions but the 4$\rm p$ $\rightarrow$ $\rm ns$ series appear to be too weak to be observed. 

%
%
%
%

%
%

\begin{table}
\begin{center}
\caption{Principal quantum number ($\rm n$), resonance
    	       energies E (eV) and quantum defect ($\mu$) from
     	       experimental measurements of Se$^+$  \cite{Esteves2010,sterling11}
   	       compared  with present theoretical estimates from the QB method. 
    	      The Rydberg series originating from the $\rm ^2P^o_{3/2}$ metastable 
	      state of Se$^{+}$  due to $\rm 4p \rightarrow nd$ transitions is tabulated. The uncertainties in
    	      the experimental energies are stated to be $\pm$0.010 eV or less. The series limits are taken from the 
      	     NIST \cite{Ralchenko2010}  tabulations and from
	      Esteves and co-workers \cite{Esteves2010,sterling11}. }
\begin{tabular}{cccc@{\ \ \ \ \ \ \ \ \ \ \ \ }cccc}
\hline\hline\\
\multicolumn{3}{c}{\bfseries Initial Se$^+$ state}&&\multicolumn{3}{c}{\bfseries Initial Se$^+$ state}\\
\multicolumn{3}{c}{$\rm \:4s^24p^3\ (^2P^{o}_{3/2})$}&&\multicolumn{3}{c}{$\rm \:4s^24p^3\ (^2P^{o}_{3/2})$}\\
\hline
\multicolumn{3}{c}{Rydberg Series}&&\multicolumn{3}{c}{Rydberg Series}\\[.02in]
\multicolumn{3}{c}{\small{$\rm \:4s^24p^2(^1D_2)\,nd $}}&&\multicolumn{3}{c}{\small{$\rm \:4s^24p^2(^1D_2)\,nd$}}\\
n			&E (eV)			&$\mu$	&	&n		&E (eV)&$\mu$\\[.02in]
\hline\\
\multicolumn{3}{c}{[Theory]}&&\multicolumn{3}{c}{[Experiment] }\\[.02in]
6			&18.256				&0.14	&	& 6		&18.268				&0.14\\
7			&18.684				&0.15	&	& 7		&18.697				&0.14\\
8			&18.957				&0.16	&	& 8		&18.972				&0.14\\
9			&19.149				&0.14	&	& 9		&19.160				&0.14\\
10			&19.280				&0.16	&	& 10		&19.296				&0.12\\
11			&19.379				&0.15	&	& 11 		&19.395				&0.10\\
12			&19.455				&0.14	&	& -		& -					&-\\
$\cdot$		&$\cdot$				&-		&	&$\cdot$	&$\cdot$				&-\\
$\infty$		&19.842$^{\dagger}$	&		&	&$\infty$	&19.853 $\pm$ 0.05$^{\ddagger}$	&\\
\hline\hline\\
$^{\dagger}$NIST  tabulations \cite{Ralchenko2010}\\
$^{\ddagger}$Esteves PhD thesis \cite{Esteves2010}\\
\end{tabular}
\end{center}
\end{table}
%
%

\begin{table}
\begin{center}
\caption{ Principal quantum number ($\rm n$), resonance
    	 	 energies E (eV) and quantum defect ($\mu$) from
     		experimental measurements of Se$^+$  \cite{Esteves2010,sterling11}
   		 compared  with present theoretical estimates from the QB method. 
    		The Rydberg series originating from the $\rm ^2P^o_{3/2}$ metastable 
		state of Se$^{+}$ due to $\rm 4p \rightarrow nd$ transitions is tabulated. 
		The uncertainties of  the experimental energies are stated
		to be $\pm$0.010 eV. The series limits are taken from the 
      		NIST \cite{Ralchenko2010}  tabulations and 
		from Esteves and co-workers \cite{Esteves2010,sterling11}. }
\begin{tabular}{cccc@{\ \ \ \ \ \ \ \ \ \ \ \ }cccc}
\hline\hline\\
\multicolumn{3}{c}{\bfseries Initial Se$^+$ state}&&\multicolumn{3}{c}{\bfseries Initial Se$^+$ state}\\
\multicolumn{3}{c}{$\rm \:4s^24p^3\ (^2P^{o}_{3/2})$}&&\multicolumn{3}{c}{$\rm \:4s^24p^3\ (^2P^{o}_{3/2})$}\\
\hline
\multicolumn{3}{c}{Rydberg Series}&&\multicolumn{3}{c}{Rydberg Series}\\[.02in]
\multicolumn{3}{c}{\small{$\rm \:4s^24p^2(^1S_0)\,nd $}}&&\multicolumn{3}{c}{\small{$\rm \:4s^24p^2(^1S_0)\,nd$}}\\
n			&E (eV)			&$\mu$	&	&n		&E (eV)&$\mu$\\[.02in]
\hline\\
\multicolumn{3}{c}{[Theory]}&&\multicolumn{3}{c}{[Experiment] }\\[.02in]
5			&19.350			&0.24	&	&5		&19.301				&0.22\\
6			&20.118			&0.23	&	&6		&20.074				&0.22\\
7			&20.556			&0.22	&	&7		&20.530				&0.19\\
8			&20.851			&0,22	&	&8		&20.815				&0.17\\
9			&21.044			&0.22	&	&9		&21.008				&0.15\\
10			&21.181			&0.22	&	&10		&21.149				&0.10\\
11			&21.283			&0.22	&	&-		&-					&-\\
12			&21.359			&0.22	&	&-		&-					&-\\
$\cdot$		&$\cdot$			&--		&	&$\cdot$	&$\cdot$				&-\\
$\infty$		&21.751$^{\dagger}$	&		&		&$\infty$	&21.703$\pm$ 0.05 $^{\dagger}$	&\\
\hline\hline\\
$^{\dagger}$NIST  tabulations \cite{Ralchenko2010}\\
$^{\ddagger}$Esteves PhD thesis \cite{Esteves2010}\\
\end{tabular}
\end{center}
\end{table}

%
%

\begin{table}
\begin{center}
\caption{ Principal quantum number ($\rm n$), resonance
    	 	 energies E (eV) and quantum defect ($\mu$) from
     		experimental measurements of Se$^+$  \cite{Esteves2010,sterling11}
   		 compared  with present theoretical estimates from the QB method. 
    		 The Rydberg series originating from the $\rm ^2P^o_{1/2}$ metastable 
		 state of Se$^{+}$ due to $\rm 4p \rightarrow nd$ transitions is tabulated. 
		 The uncertainties of the experimental energies are stated to 
		 be $\pm$0.010 eV. The series limits are taken from the 
      		NIST \cite{Ralchenko2010}  tabulations and 
		from Esteves and co-workers \cite{Esteves2010,sterling11}.}
\begin{tabular}{cccc@{\ \ \ \ \ \ \ \ \ \ \ \ }cccc}
\hline\hline\\
\multicolumn{3}{c}{\bfseries Initial Se$^+$ state}&&\multicolumn{3}{c}{\bfseries Initial Se$^+$ state}\\
\multicolumn{3}{c}{$\rm \:4s^24p^3\ (^2P^{o}_{1/2})$}&&\multicolumn{3}{c}{$\rm \:4s^24p^3\ (^2P^{o}_{1/2})$}\\
\hline
\multicolumn{3}{c}{Rydberg Series}&&\multicolumn{3}{c}{Rydberg Series}\\[.02in]
\multicolumn{3}{c}{\small{$\rm \:4s^24p^2(^1D_2)\,nd $}}&&\multicolumn{3}{c}{\small{$\rm \:4s^24p^2(^1D_2)\,nd$}}\\
n			&E (eV)			&$\mu$	&	&n		&E (eV)&$\mu$\\[.02in]
\hline\\
\multicolumn{3}{c}{[Theory]}&&\multicolumn{3}{c}{[Experiment] }\\[.02in]
6			&18.345				&0.17	&	& 6	&18.354				&0.17\\
7			&18.786				&0.16	&	& 7	&18.788				&0.17\\
8			&19.062				&0.16	&	& 8	&19.070				&0.16\\
9			&19.251				&0.16	&	& 9	&19.262				&0.14\\
10 			&19.385				&0.16	&	&10	&19.298				&0.12\\
11			&19.485				&0.16	&	&-	&-					&-	\\
12			&19.561				&0.15	&	&-	&-					&-	\\
$\cdot$		&$\cdot$				&-		&	&-	&$\cdot$				&-\\
$\infty$		&19.842$^{\dagger}$	&		&	&	&19.853 $\pm$ 0.05$^{\ddagger}$	&\\
\hline\hline\\
$^{\dagger}$NIST  tabulations \cite{Ralchenko2010}\\
$^{\ddagger}$Esteves PhD thesis \cite{Esteves2010}\\
\end{tabular}
\end{center}
\end{table}

%
%

\begin{table}
\begin{center}
\caption{ Principal quantum number ($\rm n$), resonance
    	 	 energies E (eV) and quantum defect ($\mu$) from
     		experimental measurements of Se$^+$  \cite{Esteves2010,sterling11}
   		 compared  with present theoretical estimates from the QB method. 
    		The Rydberg series originating from the $\rm ^2P^o_{1/2}$ metastable 
		state of Se$^{+}$ due to $\rm 4p \rightarrow nd$ transitions is tabulated. 
		The uncertainties in  the experimental energies are stated
		to be $\pm$0.010 eV. The series limits are taken from the 
      		NIST \cite{Ralchenko2010}   tabulations and 
		from Esteves and co-workers \cite{Esteves2010,sterling11}.}
\begin{tabular}{cccc@{\ \ \ \ \ \ \ \ \ \ \ \ }cccc}
\hline\hline\\
\multicolumn{3}{c}{\bfseries Initial Se$^+$ state}&&\multicolumn{3}{c}{\bfseries Initial Se$^+$ state}\\
\multicolumn{3}{c}{$\rm \:4s^24p^3\ (^2P^{o}_{1/2})$}&&\multicolumn{3}{c}{$\rm \:4s^24p^3\ (^2P^{o}_{1/2})$}\\
\hline
\multicolumn{3}{c}{Rydberg Series}&&\multicolumn{3}{c}{Rydberg Series}\\[.02in]
\multicolumn{3}{c}{\small{$\rm \:4s^24p^2(^1S_0)\,nd $}}&&\multicolumn{3}{c}{\small{$\rm \:4s^24p^2(^1S_0)\,nd$}}\\
n			&E (eV)			&$\mu$	&	&n		&E (eV)&$\mu$\\[.02in]
\hline\\
\multicolumn{3}{c}{[Theory]}&&\multicolumn{3}{c}{[Experiment]$^{\ddagger}$ }\\[.02in]
5			&19.469				&0.23	&	&5		&19.468				&0.17\\
6			&20.227				&0.22	&	&6		&20.202				&0.17\\
7			&20.673				&0.22	&	&7		&20.637				&0.17\\
8			&20.958				&0.22	&	&8		&20.922				&0.15\\
9			&21.151				&0.22	&	&9		&21.116				&0.11\\
10			&21.288				&0.22	&	&10		&21.238				&0.20\\
11			&21.389				&0.22	&	&-		&-					&-\\
12			&21.465				&0.22	&	&-		&-					&-\\
$\cdot$		&$\cdot$				&-		&	&$\cdot$	&$\cdot$				&-\\
$\infty$		&21.857$^{\dagger}$	&		&	&$\infty$	&21.805 $\pm$ 0.05$^{\dagger}$	&\\
\hline\hline\\
$^{\dagger}$NIST  tabulations \cite{Ralchenko2010}\\
$^{\ddagger}$Esteves PhD thesis \cite{Esteves2010}\\
\end{tabular}
\end{center}
\end{table}

%
%

\begin{table}
\begin{center}
\caption{Principal quantum number ($\rm n$), resonance
     energies E (eV) and quantum defect ($\mu$) from
     experimental measurements of Se$^+$  \cite{Esteves2010,sterling11}
    compared  with present theoretical estimates from the QB method. 
     The Rydberg series  originating
     from the $\rm ^2D^o_{5/2}$ metastable state of Se$^+$ due to $\rm 4p \rightarrow nd$
     transitions  is tabulated. 
     The uncertainties in the experimental energies are 
     stated to be $\pm$0.010 eV. The series limits are taken from the 
      NIST \cite{Ralchenko2010}  tabulations and 
      from Esteves and co-workers \cite{Esteves2010,sterling11}.}
\begin{tabular}{cccc@{\ \ \ \ \ \ \ \ \ \ }cccc}
\hline\hline\\
\multicolumn{3}{c}{\bfseries  Initial Se$^+$ state}&&\multicolumn{3}{c}{\bfseries  Initial Se$^+$ state}\\
\multicolumn{3}{c}{$\rm \:4s^24p^3\ (^2D^{o}_{5/2})$}&&\multicolumn{3}{c}{$\rm \:4s^24p^3\ (^2D^{o}_{5/2})$}\\
\hline
\multicolumn{3}{c}{Rydberg Series}&&\multicolumn{3}{c}{Rydberg Series}\\[.02in]
\multicolumn{3}{c}{\small{$\rm \:4s^24p^2(^1D_2)\,nd $}}&&\multicolumn{3}{c}{\small{$\rm \:4s^24p^2(^1D_2)\,nd$}}\\
n			&E (eV)			&$\mu$	&	&n		&E (eV)			&$\mu$\\[.02in]
\hline\\
\multicolumn{3}{c}{[Theory]}&&\multicolumn{3}{c}{[Experiment] }\\[.02in]
6			&19.496				&0.17	&	&6		&19.499				&0.17\\
7			&19.938				&0.15	&	&7		&19.933				&0.17\\
8			&20.211				&0.16	&	&8		&20.212				&0.17\\
9			&20.399				&0.16	&	&9		&20.402				&0.16\\
10                       &20.533				&0.16	&	&10		&20.537				&0.17\\
11                       &20.633				&0.16	&	&11		&20.637				&0.16\\
12                       &20.707				&0.16	&	&12		&20.712				&0.16\\
$\cdot$		&$\cdot$				&-		&	&$\cdot$	&$\cdot$		 		&-\\
$\infty$		&21.095$^{\dagger}$	&		&	&$\infty$	&21.100$\pm$ 0.01$^{\ddagger}$	&\\
\hline\hline\\
$^{\dagger}$NIST  tabulations \cite{Ralchenko2010}\\
$^{\ddagger}$Esteves PhD thesis \cite{Esteves2010}\\
\end{tabular}
\end{center}
\end{table}

%
%

\begin{table}
\begin{center}
\caption{Principal quantum number ($\rm n$), resonance
     energies E (eV) and quantum defect ($\mu$) from
     experimental measurements of Se$^+$  \cite{Esteves2010,sterling11}
    compared  with present theoretical estimates from the QB method. 
     The Rydberg series  originating
     from the $\rm ^2D^o_{3/2}$ metastable state of Se$^+$ due to $\rm 4p \rightarrow nd$
     transitions  is tabulated. 
     The uncertainties in the experimental energies are 
     stated to be $\pm$0.010 eV. The series limits are taken from the 
      NIST \cite{Ralchenko2010} tabulations and 
      the work of Esteves and co-workers \cite{Esteves2010,sterling11}.}
\begin{tabular}{cccc@{\ \ \ \ \ \ \ \ \ \ }cccc}
\\
\hline\hline\\
\multicolumn{3}{c}{\bfseries  Initial Se$^+$ state}&&\multicolumn{3}{c}{\bfseries  Initial Se$^+$ state}\\
\multicolumn{3}{c}{$\rm \:4s^24p^3\ (^2D^{o}_{3/2})$}&&\multicolumn{3}{c}{$\rm \:4s^24p^3\ (^2D^{o}_{3/2})$}\\
\hline
\multicolumn{3}{c}{Rydberg Series }&&\multicolumn{3}{c}{Rydberg Series }\\[.02in]
\multicolumn{3}{c}{\small{$\rm \:4s^24p^2(^1D_2)\,nd $}}&&\multicolumn{3}{c}{\small{$\rm \:4s^24p^2(^1D_2)\,nd$}}\\
n			&E (eV)			&$\mu$	&	&n		&E (eV)			&$\mu$\\[.02in]
\hline\\
\multicolumn{3}{c}{[Theory]}&&\multicolumn{3}{c}{[Experiment] }\\[.02in]
6			&19.573				&0.17	&	&6		&19.572				&0.18\\
7			&20.009				&0.16	&	&7		&20.009				&0.17\\
8			&20.287				&0.16	&	&8		&20.289				&0.17\\
9			&20.476				&0.16	&	&9		&20.479				&0.16\\
10                       &20.610				&0.16	&	&10		&20.613				&0.17\\
11                       &20.710				&0.16	&	&11		&20.713				&0.16\\
12                       &20.784				&0.16	&	&12		&20.788				&0.16\\
$\cdot$		&$\cdot$				&-		&	&$\cdot$	&$\cdot$		 		&-\\
$\infty$		&21.172$^{\dagger}$	&		&	&$\infty$	&21.176 $\pm$ 0.025$^{\ddagger}$	&\\
\hline\hline\\
$^{\dagger}$NIST  tabulations~\cite{Ralchenko2010}\\
$^{\ddagger}$Esteves PhD thesis~\cite{Esteves2010}\\
\end{tabular}
\end{center}
\end{table}
%
%
%

At energies  below 21.2 eV there are prominent Rydberg resonance features
 resulting from photoionization of the initial metastable Se$^+$ states 
in the experimental and theoretical cross sections, as is seen in figure ~\ref{StackThrExp}.  
Multiple overlapping Rydberg resonance series in the spectra are also clearly visible 
converging to the ${\rm 4s^24p^2~^1D_2}$ and 
${\rm 4s^24p^2~^1S_0}$ thresholds respectively of the residual Se$^{2+}$ ion.
The presence of interloping resonances disrupt the regular
Rydberg pattern of resonances,  as can be seen in the spectra for the 
$\rm ^2P^o_{1/2}$ and $\rm ^2P^o_{3/2}$ metastable states, respectively figures \ref{fig7} and  \ref{fig8}.

To obtain the resonance parameters
 a theoretical analysis with a modification  to
the widely used eigenphase derivative (QB) technique (for $jj$ -coupling) that determines resonance
parameters in electron collisions with atomic and molecular systems
\cite{keith1996,keith1998,keith1999} was used.  This technique exploits the properties 
of the R-matrix in multi-channel scattering and is an excellent method  for locating 
and determining the properties of narrow resonances. 
For this trans-iron element  our theoretical results 
(obtained for the dominant resonance features in the photoionization spectra)
are compared with the recent ALS measurements \cite{Esteves2010,sterling11}.
From the ALS measurements \cite{Esteves2010,sterling11} 
it was shown that the $\rm ^2P^o_{1/2}$ and  $\rm ^2P^o_{3/2}$  initial metastable
states of the Se$^{+}$  ion both produce two dominant Rydberg resonance series. 
Our results for these resonances are illustrated in figure \ref{fig7} and \ref{fig8}.  
 The $\rm 4s^24p^2(^1D_2)\, nd$ series converging to the $\rm ^1D_2$
series limit is indicated with solid triangles and the $\rm 4s^24p^2(^1S_0)\, nd$ 
series converging to the $\rm ^1S_0$ limit by the open triangles.  
 Furthermore, two additional autoionizing Rydberg series were identified associated with
the $\rm ^2D^o_{3/2}$  and $\rm ^2D^o_{5/2}$ metastable states, our theoretical results are shown
by the solid triangles in figures \ref{fig9} and \ref{fig10}.
 
The relationship between the principal quantum number n, 
the effective quantum number $\nu$ and the quantum defect $\mu$ 
for an ion of effective charge ${\cal Z}$ is given by 
$\nu$ = $\rm n - \mu$ where the resonance position $\epsilon_r$ can be determined from Rydberg's formula
\begin{equation}
\epsilon_r  =  \epsilon_{\infty} -  {\cal~Z}^2 / \nu^{2},
\end{equation}
 and $\epsilon_{\infty}$ is the resonance series limit \cite{Seaton1983}. 
 Tables 2 -- 5  lists the principal quantum number n,
resonance energy $\epsilon_r$ and quantum defect $\mu$ of the first few members obtained from our
work for  each of the prominent Rydberg series originating from 
the $\rm ^2P^o_{3/2}$  and $\rm ^2P^o_{1/2}$ metastable states.
The experimental results of Esteves and co-workers \cite{Esteves2010,sterling11} 
are included in Tables 2-- 5 for comparison purposes where it is seen our theoretical work 
is in satisfactory agreement with experiment for the resonance energies and quantum defects. 
Measurements taken at the Advanced Light Source synchrotron radiation facility 
 indicated  the presence of several prominent $\rm 4s^24p^2(^1D_2)\,nd$ 
resonances in this energy range originating from the $\rm ^2D^o_{5/2}$ 
and  $\rm ^2D^o_{3/2}$  initial metastable states. 
Tables 6 and 7  present  results for the  $\rm 4s^24p^2(^1D_2)\, nd$ 
dominant resonance series observed in the PI cross sections originating
from the $\rm ^2D^o_{3/2}$  and $\rm ^2D^o_{5/2}$ metastable states respectively. 
Here again our theoretical work show satisfactory agreement  with experiment 
for the resonance energies and quantum defects.

Finally, from figure \ref{fig11}, theoretical estimates indicate many of the 
 linewidths  ($\Gamma$, determined from the QB method)  for the $\rm 4s^24p^2(^1D_2)\,nd$ series
are less than a few meV, making their experimental determination challenging.
Overall, given the complexity of this trans-iron element and the wealth of 
resonance structure in the corresponding PI cross sections, 
the agreement between theory and experiment is very satisfactory.

\section{Conclusions}
State-of-the-art theoretical methods were used to investigate 
the photoionization of Se$^{+}$ ions in the energy region of 
the ground-state ionization threshold (18 eV -- 31 eV). 
Given the complexity of this trans-iron element, overall,
 satisfactory agreement is found between
 theoretical and experimental results both on the photon-energy 
 scale and on the absolute PI cross-section scale. 
The photoionization cross section exhibits a wealth of resonances.  
Prominent members of the Rydberg series are 
analyzed and compared with experiment.
Resonance positions and quantum defects for 
the 4$\rm s^2$4$\rm p^2$($\rm ^1D_2$)$\rm nd$
and 4$\rm s^2$4$\rm p^2$($\rm ^1S_0$)$\rm nd$ series converging 
to their respective series limits are presented.
The metastable state content of the Se$^+$ ion beam has been 
determined along with prominent autoionizing Rydberg 
resonances series seen in the experimental PI cross sections.  A clear pattern of  Rydberg 
resonance series is seen in the spectra of the photoionization 
cross sections disrupted by interloping resonances (see figures 4 and 5).

We point out that the strength of the present study is in 
the convergence between two different state-of-the-art 
theoretical methods and the comparison with available experimental data. 
Since no beam diagnostics were carried out in the experiments \cite{Esteves2010,sterling11} it is 
left to theory to infer the appropriate fractions. 
Such detailed comparisons, between theory and experiment, 
strengthens the validity of our results for use in astrophysical applications. 

The photoionization cross-sections from the present study are suitable to be included into
state-of-the-art photoionization modelling codes Cloudy
and XSTAR \cite{ferland98,kallman01}  that are used to numerically
simulate the thermal and ionization structure of ionized astrophysical nebulae.
In combination with our determinations of the photoionization properties of other
low-charge Se ions (to be addressed in subsequent publications), this will enable
the abundance of Se to be accurately determined in astrophysical nebulae. Such
information bears implications for the sites of origin and chemical evolution of
Se and other trans-iron elements in the Universe, as well as for studies of the
interior structure and nucleosynthesis of giant stars.
These recent modifications to the  Dirac-Atomic-R-matrix-Codes (DARC) have now 
made it feasible to systematically study other
heavy complex systems of prime interest to 
astrophysics.

\ack
C P Ballance was supported by US Department of Energy (DoE)
grants  through Auburn University.
B M McLaughlin acknowledges support by the US
National Science Foundation through a grant to ITAMP
at the Harvard-Smithsonian Center for Astrophysics.  
We thank David A Esteves for providing us 
with the  ALS experimental data and a copy of his thesis.
The computational work was carried out at the National Energy Research Scientific
Computing Center in Oakland, CA, USA and on the Tera-grid at
the National Institute for Computational Sciences (NICS) in Knoxville, TN, USA. 
\clearpage
%
%
%
%
\bibliographystyle{iopart-num}
\bibliography{seplus}
\end{document}